\documentclass[letter]{aa}

\usepackage{graphicx}
\usepackage{txfonts}
\begin{document}

   \title{Abundance of HOCO$^+$ and CO$_2$ in the outer layers
of the L1544 prestellar core}

   \author{C. Vastel
          \inst{1,2}
          \and
          C. Ceccarelli\inst{3,4}
          \and
          B. Lefloch\inst{3,4}
           \and
          R. Bachiller\inst{5}
          }

   \institute{Universit\'e de Toulouse, UPS-OMP, IRAP, Toulouse, France
              \email{charlotte.vastel@irap.omp.eu}
         \and
             CNRS, IRAP, 9 Av. Colonel Roche, BP 44346, F-31028 Toulouse Cedex 4, France
             \and
             Universit\'e de Grenoble Alpes, IPAG, F-38000 Grenoble, France
             \and
             CNRS, IPAG, F-38000 Grenoble, France
             \and
             Observatorio Astron\'omico Nacional (OAN, IGN), Calle Alfonso XII, 3, 28014 Madrid, Spain
             }


 
  \abstract
  {The L1544 prestellar core has been observed as part of the ASAI
    IRAM Large Program at 3 mm. These observations led to the
    detection of many complex molecules. In
    this Letter, we report the detection of two lines, at 85.5 GHz
    (4$_{0,4}$--3$_{0,3}$) and 106.9 GHz (5$_{0,5}$--4$_{0,4}$),
    respectively, of the protonated carbon dioxide ion, HOCO$^+$. We
    also report the tentative detection of the line at 100.4 GHz
    (5$_{0,5}$--4$_{0,4}$) of DOCO$^+$. The non-LTE analysis of the
    detected lines shows that the HOCO$^+$ emission originates in
    the external layer where non-thermal desorption of other species
    has previously been observed. Its abundance is
    $(5\pm2)\times10^{-11}$. Modelling of the chemistry involved in the
    formation and destruction of HOCO$^+$ provides a gaseous CO$_2$ abundance of $2 \times 10^{-7}$ (with respect to H$_2$) with an upper limit of  $2 \times 10^{-6}$. }


   \maketitle
%

\section{Introduction}

The protonated form of CO$_2$, HOCO$^+$ , was first identified in the
Galactic centre molecular cloud SgrB2 by Thaddeus, Gu\'elin, and Linke
(1981). The detection was then confirmed with the laboratory detection
by Bogey, Demuynck, and Destombes (1984). Minh et al. (1988) reported a
survey towards 18 molecular clouds (dark clouds and active star-forming
regions) and concluded that the HOCO$^+$ ion was only detected
towards SgrB2, and later towards SgrA (Minh et al. 1991). To date, this ion
has been detected towards the Galactic centre, several translucent and
dark clouds (Turner et al. 1999), a single low-mass Class 0 protostar
(Sakai et al. 2008), and in the prototypical protostellar bow shock
L1157-B1 (Podio et al. 2014). HOCO$^+$ was first proposed by Herbst
et al. (1977) as an indirect tracer of gas-phase CO$_2$. Through a
comparison of the abundances of HOCO$^+$ with that of HCO$^+$, the
abundance of gas-phase CO$_2$ relative to that of CO might be
constrained.
Carbon dioxide (CO$_2$) is an important constituent of interstellar
ices and has been widely detected in absorption towards infrared
bright sources with the ISO and Spitzer telescopes. However, its
formation mechanism is still not completely understood. Unfortunately,
CO$_2$ cannot be traced in the millimeter/submillimeter regime because
it lacks a permanent dipole moment, therefore it can only be sought
towards sources with a bright infrared continuum. In the solid phase,
its abundance represents about 15--50$\%$ of solid H$_2$O in quiescent 
molecular clouds, low-mass and high-mass protostars 
(Whittet et al. 1998, 2007, 2009, \"Oberg et al. 2011), and
the ice abundances seem to be the highest for the coldest sources. The
observed abundances of solid CO$_2$ in the interstellar medium are a
factor of 100 higher than in the gas phase (van Dishoeck et al. 1996;
Boonman et al. 2003a), and the formation of CO$_2$ is therefore assumed
to proceed through reactions in the ices of interstellar dust grains.
CO$_2$ is  readily  produced  in  UV  photo-processed  CO-H$_2$O
laboratory ice, with an efficiency high enough to be driven by the
cosmic-ray-induced  UV  field  in  dense  interstellar  regions
(Watanabe \& Kouchi 2002). Cosmic-ray processing of pure CO
laboratory ice has also been shown to be a viable mechanism for CO$_2$
production in the interstellar medium (Jamieson et al. 2006) and is
an interesting solution given the large abundances of pure CO ice
(Pontoppidan et al. 2003). Recently, Ioppolo et al. (2011) showed a
correlation between the formation of CO$_2$ and H$_2$O (under
laboratory conditions), which is consistent with the astronomical
observation of solid CO$_2$ in water-rich environments. 
Neill et al. (2014) performed a sensitive line survey of Sgr B2(N), and
based on the HOCO$^+$ species, they estimated a CO$_2$/CO ratio in the colder
external envelope of between 0.01-0.1, implying a CO$_2$ gas-phase
abundance of 10$^{-6}$--10$^{-5}$, consistent with the finding of Minh
et al. (1988, 1991). Their lower limit to the CO$_2$ gas-phase
abundance of 10$^{-6}$ relative to H$_2$ indicates that the CO$_2$
abundance in the gas is clearly enhanced, while most of the water is
still frozen out in ices in the Sgr B2 envelope.

As part of the IRAM-30m Large Program ASAI{ \footnote{Astrochemical
    Surveys At Iram: http://www.oan.es/asai/} (Lefloch et al. 2016 in
  prep.), we  carried out a highly sensitive, unbiased spectral
  survey of the molecular emission of the L1544 prestellar core with high spectral resolution. This source is a prototypical starless core in the Taurus molecular cloud
complex (d $\sim$ 140 pc) on the verge of the gravitational collapse
(Caselli et al. 2002 and references within). It is characterised by a
central high density (2~10$^6$ cm$^{-3}$) and a low temperature
($\sim$ 7 K). We reported the detection of many oxygen bearing complex organic 
  molecules produced through the release in the gas phase of methanol and ethene through non-thermal desorption processes (Vastel et al. 2014, hereafter Paper I). The sensitivity of these IRAM observations also led to the detection of the hyperfine structure of CH$_2$CN (Vastel et al. 2015a). In the present study we report on the
  detection of the HOCO$^+$ ion in the L1544 prestellar core, as well
  as a tentative detection of its deuterated form, DOCO$^+$.

\section{Observations and results}

The observations were performed at the IRAM-30m towards L1544
($\alpha_{2000} = 05^h04^m17.21^s, \delta_{2000} =
25\degr10\arcmin42.8\arcsec$)
using the broad-band receiver EMIR at 3mm, connected to an FTS
spectrometer in its 50 kHz resolution mode. The beam of the
observations is 29$''$ and 23$''$ at 85 and 106 GHz, respectively. 
Line intensities are expressed in units of main-beam brightness temperature (see Paper I for more details). 

The frequencies and other spectroscopic parameters of the HOCO$^+$
transitions have been retrieved from the JPL database\footnote{http://spec.jpl.nasa.gov/} from laboratory measurements by Bogey et al. (1988). Two lines of HOCO$^+$ (the 4$_{0,4}$--3$_{0,3}$ and
5$_{0,5}$--4$_{0,4}$ at 85.531 and 106.913 GHz, respectively) lie in
the frequency range covered by the ASAI survey for upper level energies lower than 30 K and were detected. Their spectra are shown in Fig. \ref{hocop_lines}. Table \ref{lines} reports
the spectroscopic parameters as well as the properties of the two
detected lines, obtained by Gaussian fitting.

For the deuterated form of HOCO$^+$, DOCO$^+$, we used the JPL
database with the spectroscopic parameters from Bogey et
al. (1986, 1988).  Three transitions lie in the frequency range for upper level energies lower than 30 K. In
Fig. \ref{docop_line} and Table \ref{lines} we present the first
tentative detection of this ion using the 5$_{0,5}$--4$_{0,4}$ transition, as well as the upper limits on the other transitions. The observed frequency (re-shifted by
the 7.2 km/s V$_{LSR}$ of L1544) is 100359.81 MHz, compared to the
computed 100359.55 $\pm$ 0.035 MHz frequency provided by the JPL
database. This is a 0.26 MHz difference and larger than the 0.035 MHz
uncertainty quoted for this transition. However, the JPL and CDMS
databases do not contain any other transition that could contaminate
this frequency, so that we tentatively assign it to DOCO$^+$. New
spectroscopic measurements and observations with a lower rms (2 mK for a 10$\sigma$ detection) 
are necessary to confirm the detection of the
DOCO$^+$ ion.


 \begin{figure}
   \centering
   \includegraphics[width=7cm]{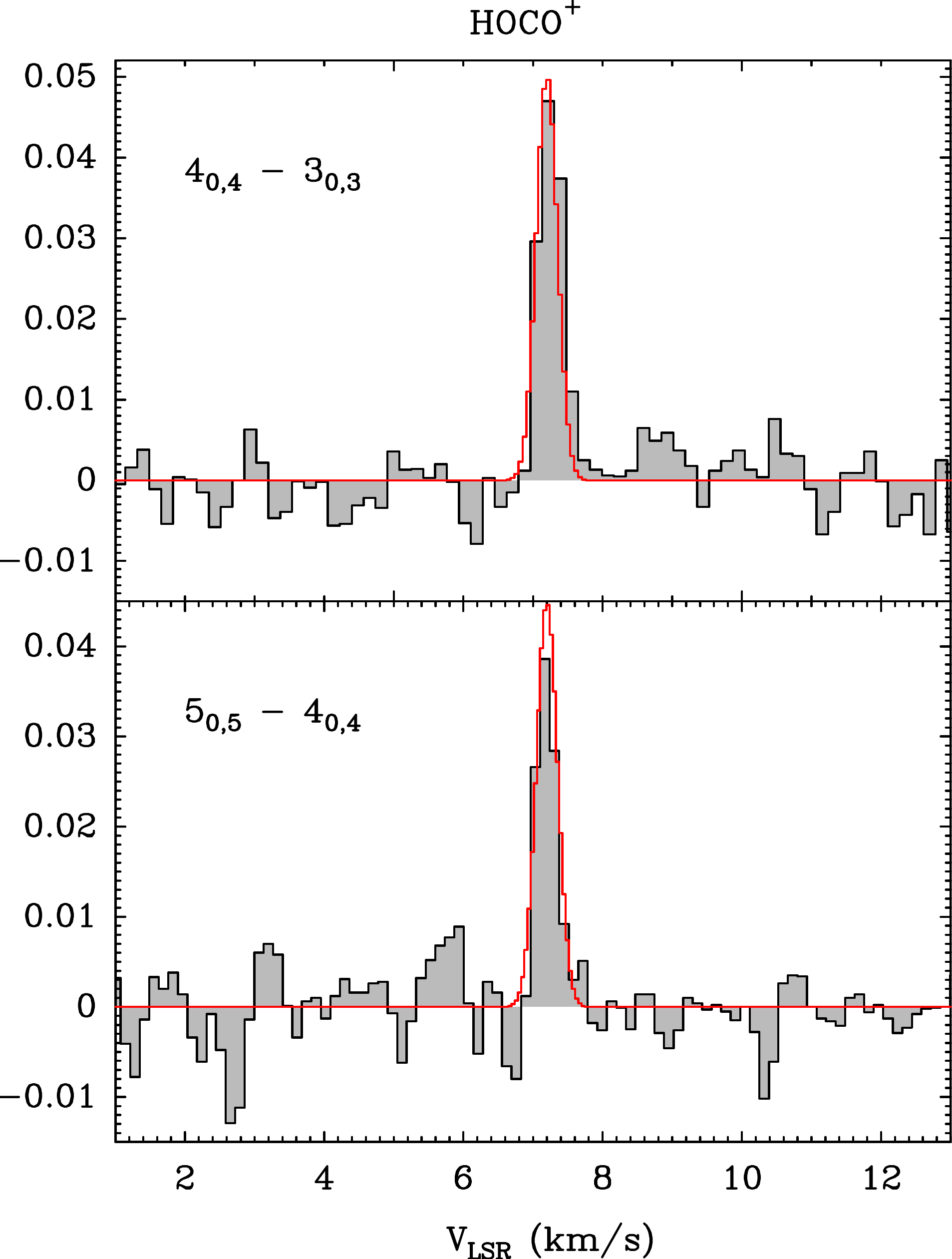}
   \caption{Detected transitions of HOCO$^+$ (T$_{mb}$). The red lines correspond to the LTE model (see text).}
   \label{hocop_lines}
 \end{figure}
 \begin{figure}
   \centering
   \includegraphics[width=7cm]{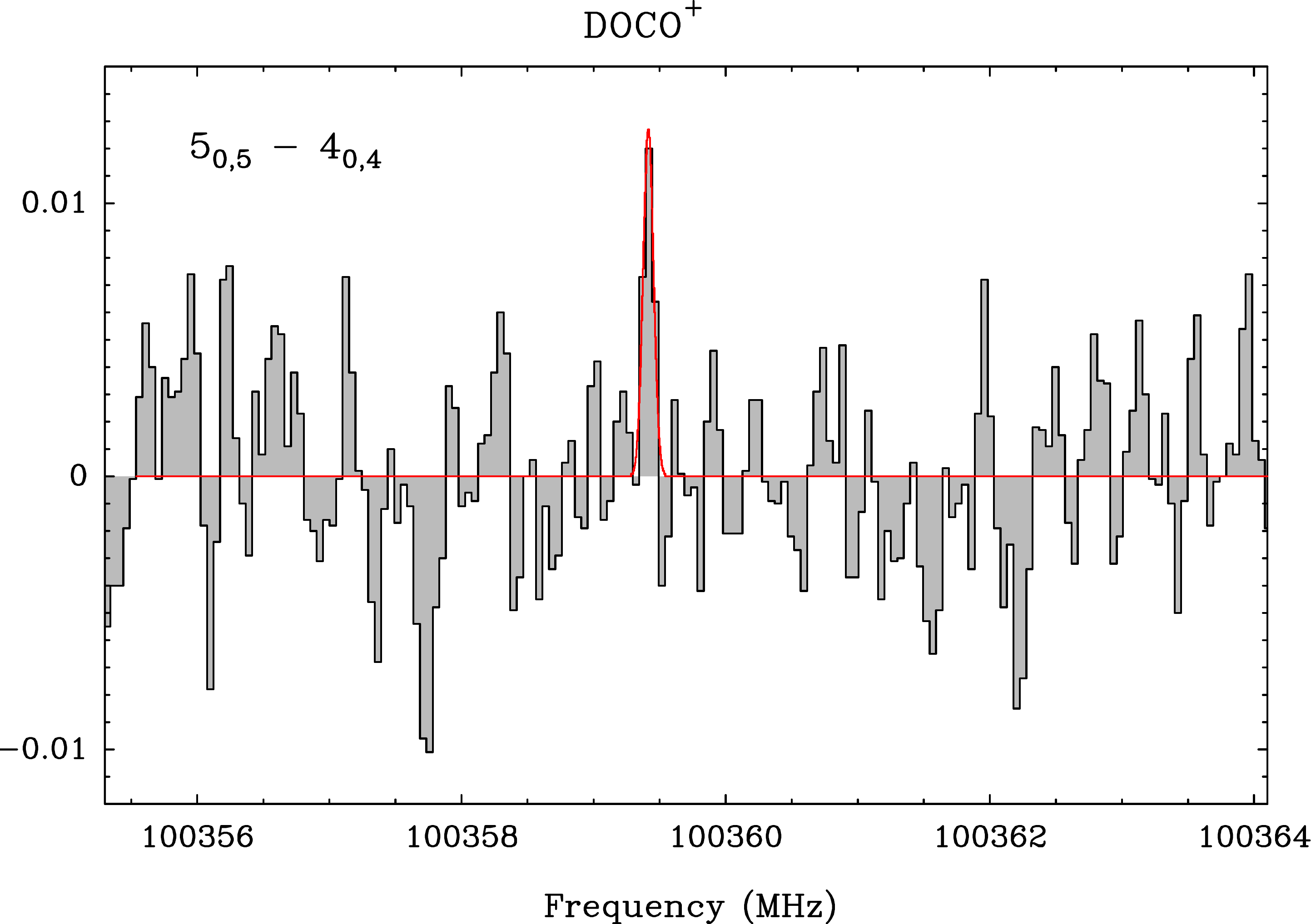}
   \caption{Tentative detection of DOCO$^+$ (T$_{mb}$). The red line corresponds to the LTE 
     model (see text). The non-detection if the 80.3 GHz transition is consistent with the derived column density, 
     taking the 5.2 mK rms into account. The modelled 80.3 GHz has the same intensity as the 100.4 GHz transition.}
   \label{docop_line}
 \end{figure}

 \begin{table*} 
 \tiny
 \caption{Properties of the observed HOCO$^+$ and
     DOCO$^+$ lines (E$_{up}$ $\le$ 30 K). The upper limits have been
     computed using the rms around the line. The errors in
     the fluxes include the calibration and statistical uncertainties
     from the Gaussian fits of the lines.\label{lines}}
\begin{tabular}{cccccccc}
Transitions  &  Frequency &        Aij        &   E$_{up}$  &  V$_{peak}$       &  FWHM              &  $\int T_{mb}dV$   &   rms\\
                   &     (MHz)      &   (s$^{-1}$) &   (K)          &  (km~s$^{-1}$)  & (km~s$^{-1}$)     &     (K~km~s$^{-1}$) & (mK)\\
\hline
                   &                    &        &          & HOCO$^+$     &  &   &\\
\hline
4$_{0,4}$--3$_{0,3}$  & 85531.51  & 1.29~10$^{-5}$ & 10.26  & 7.25 $\pm$ 0.01  & 0.43 $\pm$ 0.03  & 0.022 $\pm$ 0.004 & 3.2\\
5$_{0,5}$--4$_{0,4}$  & 106913.56 & 2.59~10$^{-5}$ &15.39  & 7.19 $\pm$ 0.02  & 0.34 $\pm$ 0.04  & 0.013 $\pm$ 0.003 & 4.6\\
\hline
&                    &                  &      & DOCO$^+$ &  \\
\hline
4$_{0,4}$--3$_{0,3}$  &  80288.79  & 1.14~10$^{-5}$ &9.63     &                             &                             & $\le$ 0.02 & 5.2\\
5$_{0,5}$--4$_{0,4}$  &  100359.55  & 2.28~10$^{-5}$ & 14.45 & 6.40 $\pm$ 0.03  & 0.28 $\pm$ 0.06  & 0.004 $\pm$ 0.001 &3.5\\
4$_{1,4}$--3$_{1,3}$  &  79776.48  & 1.05~10$^{-5}$ & 30.07     &                             &                             & $\le$ 0.02  & 4.5\\
\hline
\end{tabular}
\end{table*} 

\section{Discussion}
\subsection{Origin and abundance of HOCO$^+$ and DOCO$^+$}\label{origin}

Using the CASSIS\footnote{http://cassis.irap.omp.eu} software (Vastel et al. 2015b), 
we first performed an LTE analysis of the detected HOCO$^+$ lines of
Table \ref{lines}, where we varied the excitation temperature and
HOCO$^+$ column density. The best fit of the two lines is obtained
with a column density of $1.9~10^{11}$ cm$^{-2}$ and an
excitation temperature of 8.5 K (see Fig. \ref{hocop_lines}).  As a
second step, we performed a non-LTE modelling using the LVG code by
Ceccarelli et al. (2003) and the collision rates between HOCO$^+$ and
helium computed by Hammami et al. (2007), scaled by a factor of
1.37 to take the difference in mass of H$_2$ into account. The
collisional coefficients and spectroscopic data were
retrieved from the BASECOL database\footnote{http://basecol.obspm.fr/}.
The non-LTE LVG analysis shows a degeneracy in the density-temperature
space. In order to lift this, we used the same method as in Paper I: we plotted the $\chi^2$ contours superimposed on the
density-temperature curve of the physical structure of L1544 as
derived by Caselli et al. (2012). The result is shown in
Fig. \ref{hocop_chi2}. The two sets, LVG analysis and physical
structure, cross for a diameter of the HOCO$^+$ emitting region equal
to about 100$''$ (the emission is extended), and a HOCO$^+$
column density of $2.5~10^{11}$ cm$^{-2}$. This situation is
similar to what has been observed for the methanol emission,
which we described in Paper I.  For both methanol and HOCO$^+$, the
emission originates in the outer layers of L1544 rather than in its dense
inner core. The H$_2$ column density of this outer layer is $\sim 5~10^{21}$ cm$^{-2}$
(see Paper I), which therefore yields a HOCO$^+$ abundance of
$(5\pm2)~10^{-11}$ with respect to H$_2$.

 \begin{figure}
   \centering
   \includegraphics[angle=-90,width=7cm]{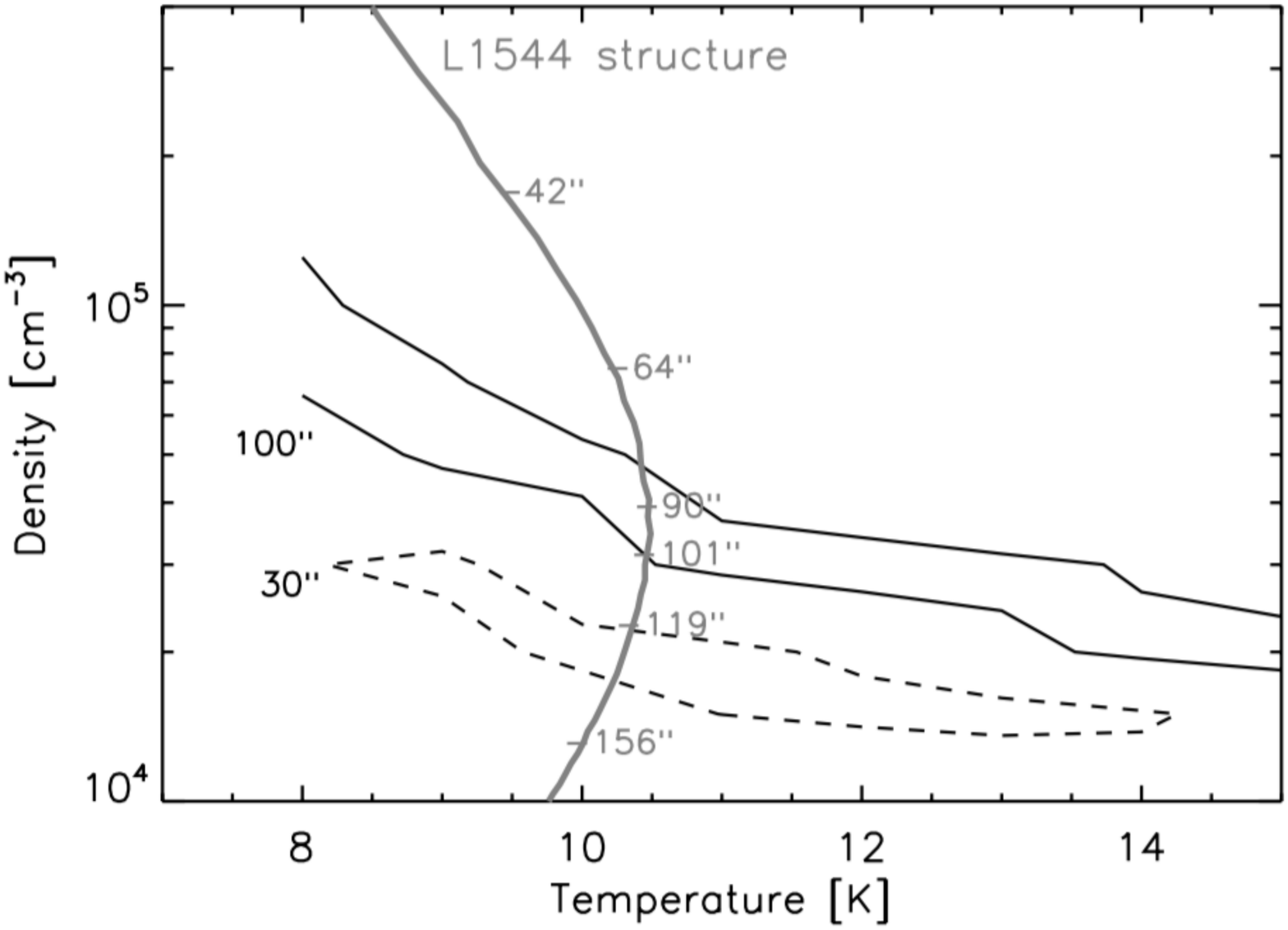}
   \caption{$\chi^2$ contour plot of the density versus temperature
     obtained by the non-LTE LVG analysis of the two detected HOCO$^+$
     lines for two different source sizes and HOCO$^+$ column
     densities: 30$''$ and $6~10^{11}$ cm$^{-2}$ (dashed line), 
     100$''$ and $2.5~10^{11}$ cm$^{-2}$ (solid line). The thick grey line shows the
     density-temperature structure of L1544 as derived by Caselli et
     al. (2012). The ticks along this line show the distance from
     the L1544 core center, in angular diameter. }
   \label{hocop_chi2}
 \end{figure}

For DOCO$^+$ we assumed the same
 excitation temperature as HOCO$^+$, namely 8.5 K, and evaluated its
 column density to be $\sim5~10^{10}$ cm$^{-2}$. Therefore, the
 upper limit on the DOCO$^+$ abundance is $\sim 10^{-11}$ and the deuteration
 ratio DOCO$^+$/HOCO$^+$ is lower than $\sim 25\%$. We note that the upper limits on the 
 80.3 and 79.8 GHz DOCO$^+$ transitions are compatible with this estimate.

\subsection{Chemical modelling and CO$_2$ gaseous abundance}

The formation of the HOCO$^+$ ion is linked to HCO$^+$ , and we can consider two main formation routes :
$\rm OH + CO \rightarrow CO_2$ followed by  
\setlength{\belowdisplayskip}{0pt} \setlength{\belowdisplayshortskip}{0pt}
\setlength{\abovedisplayskip}{0pt} \setlength{\abovedisplayshortskip}{0pt}
\begin{equation}
\rm CO_2 +H_3^+ \xrightarrow{k_1} HOCO^+ + H_2
\end{equation}
$\rm CO + H_3^+ \xrightarrow{k_2} HCO^+ + H_2 $
followed by
\begin{equation}
\rm HCO^+ + OH \xrightarrow{k_3} HOCO^+ + H
\end{equation}
The other relevant formation and destruction rates for HOCO$^+$ and HCO$^+$ are\\
$\rm HOCO^+ + CO \xrightarrow{k_4} HCO^+ + CO_2$\\
$\rm HCO^+ + e^- \xrightarrow{k_5} H + CO$\\
$\rm HCO^+ + H_2O \xrightarrow{k_6} H_3O^+ + CO$\\
$\rm HOCO^+ + e^- \xrightarrow{k_7} H + O + CO$\\
$\rm HOCO^+ + H_2O \xrightarrow{k_8} H_3O^+ + CO_2$ \\

At steady state, we can therefore express the [HOCO$^+$]/[HCO$^+$] ratio as the following:\\
\begin{equation}
\resizebox{0.9\hsize}{!}{$\frac{[HOCO^+]}{[HCO^+]} = \frac{k_1[H_3^+][CO_2]+k_3[HCO^+][OH]}{k_4[CO]+k_7[e^-]+k_8[H_2O]}\frac{k_5[e^-]+k_6[H_2O]+k_3[OH]}{k_2[CO][H_3^+]+k_4[HOCO+][CO]}$}
\end{equation}
HOCO$^+$ was first proposed by \citet{herbst1977} as an indirect tracer of gas-phase CO$_2$ and was more recently used for this purpose by \citet{sakai2008} and \citet{neill2014}. However, they neglected reaction 2 between HCO$^+$ and OH and obtained a simple relation between [HOCO$^+$]/[HCO$^+$] and [CO$_2$]/[CO] at steady state. The reaction rates (at 10 K) relevant to the formation and destruction of HOCO$^+$ and HCO$^+$ using the KIDA network\footnote{http:\slash\slash kida.obs.u-bordeaux1.fr} are the following (in cm$^3$~s$^{-1}$): k$_1$=1.90~10$^{-9}$, k$_2$=2.47~10$^{-9}$, k$_3$=5.56~10$^{-9}$, k$_4$=1.03~10$^{-9}$, k$_5$=2.93~10$^{-6}$, k$_6$=1.23~10$^{-8}$, k$_7$=7.14~10$^{-6}$, and k$_8$=1.14~10$^{-8}$. Both reactions 1 and 2 must be taken into account, and the latter cannot be neglected, leading to an indirect estimate of the [CO$_2$]/[CO] ratio from [HOCO$^+$]/[HCO$^+$].
In the centre of the prestellar core region, CO is highly depleted (Caselli et al. 1999) from the gas phase and is unlikely to produce HCO$^+$ and therefore HOCO$^+$. As a consequence, the HOCO$^+$ ion might be visible in an external layer, where CO is released from the grain surfaces in the gas phase. HOCO$^+$ depends on the cosmic ray ionisation
rate $\zeta_{CR}$, which governs the H$_3^+$ and HCO$^+$ abundances, and on the
CO$_2$ gaseous abundance when reaction (1) is more
efficient than reaction (2). We considered the observations by
Vastel et al. (2006), who measured a HCO$^+$ column density of
$4~10^{13}$ cm$^{-2}$ in the external layer of L1544. If the bulk of
the HCO$^+$ emission arises from the outer layer, whose H$_2$ column density
is $5~10^{21}$ cm$^{-2}$ (see Sect. \ref{origin}), this yields a HCO$^+$
abundance equal to $\sim8~10^{-9}$. Using the measured abundance of both HCO$^+$ and HOCO$^+$ , we can now
constrain the gaseous CO$_2$ abundance. To this end, we ran the
Nahoon gas-phase chemical model (Wakelam et al. 2015) and compared the
predicted and measured abundance of HOCO$^+$ and HCO$^+$.

 Nahoon computes the chemical evolution of a species as a function of
 time for a fixed temperature and density. The chemical network
 kida.uva.2014 contains 6992 unique chemical reactions and in total
 7506 rate coefficients, and these reactions involve 489 different
 species. We built upon the model described in Paper I, that
is, we considered
 a two-step model where in the first step we let the chemical
 composition reach steady state, and in the second step we
 injected a variable amount of CO$_2$  in the
gas phase and 
 injected methanol ($6~10^{-9}$ with respect to H atoms) and
 ethene ($5~10^{-9}$), as in Paper I, to reproduce the detected
 COMs described there. The underlying
 hypothesis is that the gaseous CO$_2$ is injected by the non-thermal
 desorption of iced CO$_2$, as in the case of methanol (Paper I). 
 
 For the first step model, we adopted an elemental gaseous carbon
 abundance of $5\times10^{-5}$ and a C/O ratio equal
 to 0.5, cosmic ionisation rate $\zeta_{CR}$ of $3~10^{-17}$
 s$^{-1}$, H density of $2\times10^4$ cm$^{-3}$, temperature of 10 K
 and A$_v$ = 10 mag, as used in Paper I. The steady abundances of
 HOCO$^+$ and HCO$^+$ are $4~10^{-11}$ and $10^{-8}$,
 respectively; this is fully consistent with the observed ones. We note that 
 a higher C/O ratio of unity will decrease to about $5~10^{-12}$ for 
 HOCO$^+$ and show no substantial change for HCO$^+$. 
We used this two-step model to derive an upper limit to
 the quantity of injected CO$_2$, taking the error bars on the observed HOCO$^+$ into account.  To stay below the upper value of the HOCO$^+$ abundance (7~10$^{-11}$), the quantity of injected CO$_2$ in the gas-phase (through the second step) must be $\leq 2~10^{-6}$  (with respect to H$_2$), namely $\sim 4\%$ of gaseous CO. We note that no CO$_2$ injection in the gas phase is compatible with our HOCO$^+$ observed abundance of 5~10$^{-11}$, with a abundance  (w.r.t H$_2$) from the gas-phase chemical modelling of $2~10^{-7}$. This value is consistent with the tentative gas-phase abundance average over the line of sight of four deeply embedded massive young stars, (van Dishoeck et al. 1996), the value found in the direction of Orion-IRc2/BN (Boonman et al. 2003a) as well as the average value found towards 8 massive protostars (Boonman et al. 2003b). The inferred abundance is also lower than the solid-state abundances of $\sim$ 1--3~10$^{-6}$ (Gerakines et al. 1999). Our study leads to the first indirect estimate of the CO$_2$ presence in the earliest phases of low-mass star-forming regions, with no confusion between hot, warm and cold gas.

Putting together the previous observations in L1544 provides us with a
   rough estimate of the composition of the grain mantles of a
   prestellar core, which can only be guessed by the indirect process
   of non-thermal desorption. It turns out that the mantles should
   contain methanol, ethene and water, while it is
   not clear that frozen CO$_2$ is present. Indeed, to convert the quoted abundances of
   gaseous into frozen species one has to take into account the
   relevant efficiency of the non-thermal desorption. Three mechanisms are invoked in the literature: FUV
   photo-desorption, cosmic-rays and chemical desorption. The latter seems to be very inefficient for the methanol
   release, based on laboratory experiments (Minissale et al. 2016). Cosmic-rays are suspected to be responsible for the re-injection into the gas phase of CO and other species with low binding energies via hot spot heating $\sim1000$ K (Leger et al. 1985; Shen et al. 2004). However, this mechanism is not efficient in releasing more strongly bonded molecules. Another possibility is that frozen molecules are injected into the gas phase by the FUV photons created by the cosmic-ray interaction with H$_2$ molecules.  Very recent experimental
   works show for the FUV photo-desorption that methanol has a very low yield, $\leq 10^{-5}$
   molecules/photon (Bertin et al. 2016),
   and that the FUV photons impinging on a methanol ice desorb
   fragments of the molecule, including CH$_3$O, with the same
   efficiency. Our value for the methanol abundance and our upper limit on the CH$_3$O abundance 
   (respectively 6~10$^{-9}$ and $\leq$ 1.5~10$^{-10}$; Paper I) differ from these
   experiments, however. In laboratory experiments, CO$_2$ is desorbed from a pure
   CO$_2$ ice FUV illuminated with a yield of $\sim 10^{-4}$
   molecules/photon (Mart\'in-Dom\'enech et al. 2015). Assuming the FUV
   photo-desorption is the dominating mechanism and neglecting the
   dependence of the photo-desorption yield on the FUV illuminating
   field, our observations would then indicate a relative composition
   of the ice as follows:  iced-CO$_2$ $\leq$ 20 times iced methanol.

\subsection{Deuteration of HOCO$^+$}

A high deuterium fractionation has been detected in the L1544 prestellar core where CO is depleted from the gas phase (Caselli et al. 1999). Caselli et al. (2003) detected a strong H$_2$D$^+$ emission, and Vastel et al. (2006) found that this ion is extended over the whole core, concluding that H$_2$D$^+$ is one of the main molecular ions in the central dense and cold region. Vastel et al. (2006) also detected a correlation between the  H$_2$D$^+$ abundance and the ratio of DCO$^+$/HCO$^+$ and N$_2$D$^+$/N$_2$H$^+$  as well as the depletion factor, because H$_2$D$^+$ is the main driver for deuteration when CO is depleted onto the grain surfaces. L1544 is the site for many first detections in prestellar cores. For example, H$_2$O has been detected with the Herschel/HIFI instrument (Caselli et al. 2012), but deuterated water
is unfortunately not detected in this source, although an upper limit on the D/H ratio has been determined from deep observations with the APEX telescope (Qu\'enard et al. 2016).
The deuterated form of HOCO$^+$ is likely produced through the high abundance of the H$_2$D$^+$ ion that appears extended in the L1544 core (Vastel et al. 2006), coupled with the non-thermal desorption of CO$_2$ in the external layer. Both H$_2$D$^+$ and DCO$^+$ detected in this core are expected to contribute to the production of the DOCO$^+$ , although no deuterated chemical network is available at the moment. The DCO$^+$/HCO$^+$ ratio in the external envelope is about 1$\%$, which is much lower than the upper limit of 25$\%$ for the DOCO$^+$/HOCO$^+$ ratio using our tentative detection. \\

\section{Conclusions}
We detected two transitions of the HOCO$^+$ ion at 85.5 and 106.9 GHz and obtained a tentative detection of a low energy level transition of its deuterated form at 100.3 GHz. Based on the HOCO$^+$ detections, we were able to estimate the column density and abundance, which was compared to a chemical modelling, constraining the properties and formation of HOCO+ at the early stages towards collapse. The result shows that our detection is compatible with an emission in an external layer, 
with a CO$_2$/H$_2$ abundance of  2~10$^{-7}$ and an upper limit of  $2 \times 10^{-6}$.

      

%

\begin{thebibliography}{}
\bibitem[Bertin et al. (2016)]{bertin2016}
Bertin, M., Romanzin, C., Doronin, M., et al. 2016, \apj, 817, L12
\bibitem[Bogey  et al. (1984)]{bogey1984} 
Bogey, M., Demuynck, C., Destombes, J.L., 1984, \aap, 138, L11
\bibitem[Bogey  et al. (1986)]{bogey1986} 
Bogey, M., Demuynck, C., Destombes, J.L., 1986, \jcp, 84, 10
\bibitem[Bogey  et al. (1988)]{bogey1988} 
Bogey, M., Demuynck, C., Destombes, J.L., Krupnov, A., 1988, Journal of Molecular Structure 190, 465
\bibitem[Boonman et al. (2003a)]{boonman2003a} 
Boonman, A.M.S., van Dishoeck, E.F., Lahuis, F., et al., 2003a, \aap, 399, 1047
\bibitem[Boonman et al. (2003b)]{boonman2003b} 
Boonman, A.M.S., van Dishoeck, E.F., Lahuis, F., Doty, S., 2003b, \aap, 399, 1063
\bibitem[Caselli et al. (1999)]{caselli1999} 
Caselli, P., Walmsley, C.M., Tafalla, M., et al., 1999, \apj, 523, L165
\bibitem[Caselli et al.(2002)]{caselli2002}
Caselli P., Walmsley C. M., Zucconi A. et al. 2002, \apj, 565, 331
\bibitem[Caselli et al.(2003)]{caselli2003}
Caselli P., van der Tak, F. F. S., Ceccarelli, C., Bacmann, A., 2003, \aap, 403, L37
\bibitem[Caselli et al. (2012)]{caselli2012} 
Caselli, P., Keto, E., Bergin, E. A., et al. 2012, \apjl, 759, L37 
\bibitem[Ceccarelli et al. (2003)]{ceccarelli2003}
Ceccarelli, C., Maret, S., Tielens, A. G. G. M., et al., 2003, \aap, 410, 587
\bibitem[Gerakines et al. (1999)]{gerakines1999} 
Gerakines, P. A., Whittet, D. C. B., Ehrenfreund, P., et al., 1999, \apj, 522, 357
\bibitem[Hammami et al. (2007)]{hammami2007}
Hammami, K., Lique, F., Ja{\"i}dane, N., et al., 2007, \aap, 462, 789 
\bibitem[Herbst et al. (1977)]{herbst1977}
Herbst, E., Green, S., Thaddeus, P., Klemperer, W., 1977, \apj, 215, 503
\bibitem[Ioppolo et al. (2011)]{ioppolo2011}
Ioppolo, S., van Boheemen, Y., Cuppen, H.M., et al., 2011, \mnras, 413, 2281 
\bibitem[Jamieson et al. (2006)]{jamieson2006}
Jamieson, C.S., Mebel, A.M., Kaiser, R.I., 2006, \apjs, 163, 184
\bibitem[Leger et al. (1985)]{leger1985}
Leger, A., Jura, M., Omont, A., 1985, \aap, 144, 147
\bibitem[Mart\'in-Dom\'enech et al. (2015)]{martindomenech2015}
Mart\'in-Dom\'enech, R., Manzano-Santamar\'ia, J., Mu\~noz Caro, G. M., et al., 2015, \aap, 584, 14
\bibitem[Minh et al. (1988)]{minh1988}
Minh, Y. C., Irvine, W. M., Ziurys, L. M, 1988, \apj, 334, 175
\bibitem[Minh et al. (1991)]{minh1991}
Minh, Y. C., Brewer, M. K., Irvine, W. M. et al., 1991, \aap, 244, 470
\bibitem[Minissale et al. (2016)]{Minissale et al. 2016}
Minissale, M, Moudens, A, Baouche, S. et al., 2016, accepted in MNRAS
\bibitem[Neill et al. (2014)]{neill2014}
Neill, J.L., Bergin, E.A., Lis, D.C., et al., 2014, \apj, 789, 8
\bibitem[Oberg et al. (2011)]{oberg2011}
\"Oberg, K.I., Boogert, A.C.A., Pontoppidan, K.M., et al., 2011, \apj, 740, 109
\bibitem[Podio et al. (2014)]{podio2014}
Podio, L., Lefloch, B., Ceccarelli, C., et al., 2014, \aap, 565, 64
\bibitem[Pontoppidan et al. (2003)]{pontoppidan2003}
Pontoppidan, K. M., Fraser, H. J., Dartois, E., et al., 2003, \aap, 408, 981
\bibitem[Qu\'enard et al. (2016)]{quenard2016}
Qu\'enard, D., Taquet, V., Vastel, C., et al., 2016, \aap, 585, 36
\bibitem[Sakai et al. (2008)]{sakai2008}
Sakai, N., Sakai, T., Aikawa, Y., Yamamoto, S., 2008, \apjl, 675, L89
\bibitem[Shen et al. (2004)]{shen2004}
Shen, C. J., Greenberg, J. M., Schutte, W. A., van Dishoeck, E. F., 2004, \aap, 415, 203
\bibitem[Thaddeus et al. (1981)]{thaddeus1981}
Thaddeus, P., Guelin, M., Linke, R.~A., 1981, \apjl, 246, L41
\bibitem[Turner et al. (1999)]{turner1999}
Turner, B.E., Terzieva, R., Herbst, E., 1999, \apj, 518, 699
\bibitem[van Dishoeck et al. (1996)]{vandishoeck1996}
van Dishoeck, E.F., Helmich, F.P., de Graauw, T., et al., 1996, \aap, 315, L349
\bibitem[Vastel et al. (2006)]{vastel2006}
Vastel, C., Caselli, P., Ceccarelli, C., et al. 2006, \apj, 645, 1198
\bibitem[Vastel et al. (2014)]{vastel2014}
Vastel, C., Ceccarelli, C., Lefloch, B., Bachiller, R., 2014, \apj, 795, L2
\bibitem[Vastel et al. (2015)]{vastel2015a}
Vastel, C., Yamamoto, S., Lefloch, B., Bachiller, R., 2015a, \aap, 582, L3
\bibitem[Vastel et al. (2015)]{vastel2015b}
Vastel, C., Bottinelli, S., Caux, E., et al., 2015b, SF2A-2015: Proceedings of the Annual meeting of the French Society of Astronomy and Astrophysics, pp.313-316
\bibitem[Wakelam et al. (2015)]{wakelam2015}
Wakelam, V., Loison, J.-C., Herbst, E., et al., 2015, \apjs, 217, 20
\bibitem[Watanabe \& Kouchi (2002)]{watanabe2002}
Watanabe, N., Kouchi, A., 2002, \apj, 567, 651
\bibitem[Whittet et al. (1998)]{whittet1998}
Whittet, D. C. B., Gerakines, P. A., Tielens, A. G. G. M., et al., 1998, \apj, 498, L159
\bibitem[Whittet et al. (2007)]{whittet2007}
Whittet, D. C. B., Shenoy, S. S., Bergin, E. A, et al., 2007, \apj, 655, 332
\bibitem[Whittet et al. (2009)]{whittet2009}
Whittet, D. C. B., Cook, A. M., Chiar, J. E., et al., 2009, \apj, 695, 94

\end{thebibliography}
%

{}

\end{document}